\documentclass[iicol]{sn-jnl}
\usepackage{graphicx}%
\usepackage{multirow}%
\usepackage{amsmath,amssymb,amsfonts}%
\usepackage{amsthm}%
\usepackage{mathrsfs}%
\usepackage[title]{appendix}%
\usepackage{xcolor}%
\usepackage{textcomp}%
\usepackage{manyfoot}%
\usepackage{booktabs}%
\usepackage{algorithm}%
\usepackage{algorithmicx}%
\usepackage{algpseudocode}%
\usepackage{listings}%
\usepackage{bm}
\usepackage{hyperref}
\newcommand{\bd}{\bm}
\begin{document}

\title{Plasmon-sound hybridization in ionic crystals}

\author{\fnm{Jakob} \sur{Rappolt}}\email{rappolt@itp.uni-frankfurt.de}

\author*{\fnm{Andreas} \sur{R\"{u}ckriegel}*}\email{rueckriegel@itp.uni-frankfurt.de}

\author{\fnm{Peter} \sur{Kopietz}}\email{kopietz@itp.uni-frankfurt.de}

\affil{\orgdiv{Institut f\"{u}r Theoretische Physik}, \orgname{Universit\"{a}t
 Frankfurt}, \orgaddress{\street{Max-von-Laue Stra{\ss}e 1}, \city{Frankfurt}, \postcode{60438}, \state{Hesse}, \country{Germany}}}

%\date{\today}
\date{January 23, 2025}

\abstract{We study the hybridization between plasmons, phonons, and electronic sound in ionic crystals using
the Debye model, 
where the ionic background is modeled as a homogeneous, isotropic, elastic medium. 
We explicitly obtain the energies and the damping of the hybrid plasmon-sound modes in the hydrodynamic regime and calculate the corresponding dynamic structure factor. 
We find that with increasing viscosity a plasmon-like mode quickly decays into a broad, incoherent background,  while a phonon-like mode with linear dispersion remains rather sharp.  
The quantitative behavior of the hybridized collective modes depends on the ratio of the electronic and the ionic plasma frequencies.
We also show that the direct Coulomb interaction between the ions is essential to obtain a collective sound mode with linear dispersion.}

\maketitle

\section{Introduction}

\label{sec:introduction}

Long-range Coulomb interactions can qualitatively modify the energy of collective modes in condensed matter. 
For example, 
the Coulomb interaction pushes the energy of the collective phase mode (Bogoliubov-Anderson mode) in a superconductor up to the electronic plasma frequency \cite{Anderson63,Schrieffer64}. 
On the other hand, 
the gapless nature of the collective phase mode in an electronic incommensurate charge-density wave has recently been shown to be preserved in the presence of Coulomb interactions \cite{Hansen23}.  
An important insight of Ref.~\cite{Hansen23} is that 
for the calculation of the collective modes in charge-density wave systems the direct Coulomb interaction 
$ \mathcal{ U }_{ i i } $ 
between positively charged ions 
[defined in Eq.~\eqref{eq:Uii} below]
cannot be neglected 
because it contributes to the balance between positive and negative charges
that is essential for a correct description of screening. 
Unfortunately, 
in classic textbooks the interaction 
$ \mathcal{ U }_{ i i } $  
is either completely ignored \cite{Pines89} or taken into account only implicitly by replacing the long-range Coulomb interaction by an effective screened interaction \cite{Fetter71}. 
In this work we show that in the Debye model, 
where the ionic background is treated as a homogeneous, isotropic, elastic medium \cite{Fetter71}, a proper microscopic treatment of
the direct Coulomb interaction between ions is essential to obtain 
the correct behavior of
collective plasmon-sound modes.

Early studies of the
hybridization between plasmons and optical phonons in semiconductors 
have already been carried out in the 1960s \cite{Varga65,Mooradian66}.
Driven by experimental and methodological advances,
plasmon-phonon mixing in various materials continues to be an active field of research.
For example, plasmon-phonon hybridization in doped semiconductors has recently been studied using  state-of-the-art first principles methods \cite{Lihm24}.  
Moreover, 
hybrid plasmon-phonon modes have also been investigated in graphene \cite{Hwang10,Sarma20,Jalalvandi21}
and in the cuprate superconductors \cite{Falter94,Falter02,Bauer09,Hepting22}.
Here we will focus on  ionic crystals
where the Coulomb interaction between positively charged ions and conduction electrons gives rise to a hybridization between the collective oscillations of the lattice (phonons) and the electronic charge (electronic sound and plasmons). 
We will mainly consider  the 
hydrodynamic regime where collisions are sufficiently frequent to generate a state of local thermal equilibrium.
Due to new experimental developments \cite{Polini20} the hydrodynamic regime of coupled electron-phonon systems has recently attracted renewed attention \cite{Levchenko20,Huang21}.
%The dynamics of a coupled fluid of electrons and acoustic phonons in the hydrodynamic regime
%where collisions 
%has been discussed long time ago \cite{Steinberg58,Gurzhi72}. 
%Here, 
%we are interested in the effect of long-range Coulomb interactions on %collective charge modes in the hydrodynamic regime of an ionic %crystal. 
Our main result is that
the existence of a sound mode with linear dispersion is immune to long-range Coulomb interactions, while the quantitative behavior of the 
hybridized plasmon-sound modes is sensitive to the ratio of the electronic and the ionic plasma frequencies.

\section{Debye model for an ionic crystal}

\label{sec:model}

The Hamiltonian of the Debye model for an ionic crystal is \cite{Fetter71}
\begin{equation}
\mathcal{H} = 
\sum_{\bd{k} } \epsilon_{\bd{k}} 
c^{\dagger}_{\bd{k} } c_{\bd{k} } +
\sum_{\bd{q}} \omega_{\bd{q}} 
b^{\dagger}_{\bd{q}} b_{\bd{q}}
+ \frac{ 1 }{ 2 \mathcal{ V } }  
\sum_{ \bd{q}  \neq 0 }  
f_{ \bd{q} } \rho_{ - \bd{q} } \rho_{ \bd{q} } , 
\label{eq:HDebye}
\end{equation}
where $ c_{ \bd{k} } $ annihilates an electron with momentum $ {\bd{k}} $ and energy 
$ \epsilon_{ \bd{k} } = k^2 / ( 2 m ) $, 
while $ b_{ \bd{q} } $ annihilates a longitudinal acoustic phonon with momentum $ \bd{q} $ and energy 
$ \omega_{ \bd{q} } = c q $. 
Here
$ k = | \bd{k} | $, $ q = | \bd{q} |$, and we ignore the spin degree of freedom.
To regularize the momentum sums we assume that the system is confined to a finite volume 
$ \mathcal{ V } $ 
and eventually consider the limit 
$ \mathcal{ V } \to \infty $, 
where the Fourier transform of the bare Coulomb interaction is given by
\begin{equation}
f_{ \bd{q} } = \frac{ 4 \pi e^2 }{ q^2 } .
\end{equation}
The Fourier components 
$ \rho_{ \bd{q} }$ 
of the total density operator have an electronic and an ionic component,
$ \rho_{ \bd{q} } 
= \rho_{ \bd{q} }^e - \rho_{ \bd{q} }^i $,
where the electronic part is 
\begin{equation}
\rho_{ \bd{q} }^e = \sum_{ \bd{k} } 
c^\dagger_{\bd{k} } c_{ \bd{k} + \bd{q} } ,
\end{equation}
while the ionic part is \cite{Fetter71}
\begin{equation}
\rho_{ \bd{q} }^i 
=
- z \sqrt{ N } 
\frac{ q }{ \sqrt{ 2 M \omega_{ \bd{q} } } }
\left( 
b_{ \bd{q} } + b^\dagger_{ - \bd{q} } 
\right) 
= 
- \sqrt{ \mathcal{V} } 
\lambda_{ \bd{q} } X_{ \bd{q} } ,
\end{equation}
with
\begin{equation}
\lambda_{ \bd{q} } = 
z \sqrt{ \frac{ n_i }{ M } } q,
\label{eq:gammaqdef}
\end{equation}
and 
\begin{equation}
X_{\bd{q}} = 
\frac{ 
b_{ \bd{q} } + b^\dagger_{ - \bd{q} } 
}{
\sqrt{ 2 \omega_{ \bd{q} } } } .
\label{eq:Xdef}
\end{equation}
Here, 
$ M $ is the ionic mass and we assume that each of the $ N $ ions contributes $z$ conduction electrons, 
so that the density of the ions is 
$ n_i = N / \mathcal{V} $, 
while the electronic density is 
$ n_e = z n_i $.
To explicitly show the electron-phonon interaction hidden in the Debye Hamiltonian \eqref{eq:HDebye}, 
we rewrite it in the following form, 
\begin{align}
\mathcal{H} 
= {} & 
\sum_{ \bd{k} } \epsilon_{ \bd{k} } 
c^\dagger_{ \bd{k} } c_{ \bd{k} } +
\sum_{ \bd{q} } \omega_{ \bd{q} }  
b^\dagger_{ \bd{q} } b_{ \bd{q} } 
+ \frac{ 1 }{2 \mathcal{V} }  
\sum_{ \bd{q} \neq 0 }  f_{ \bd{q} } 
\rho^e_{ - \bd{q} } \rho^e_{ \bd{q} }
\nonumber\\
& + 
\frac{ 1 }{ \sqrt{ \mathcal{V} } }  
\sum_{ \bd{k} , \bd{q} \neq 0 }  g_{ \bd{q} } 
c^\dagger_{ \bd{k} + \bd{q} } c_{ \bd{k} } 
\left( 
b_{ \bd{q} } + b^\dagger_{ - \bd{q} }
\right)
\nonumber\\
& + 
\frac{ 1 }{ 2 } \sum_{ \bd{q} \neq 0 }  
\frac{ \Omega_i^2 }{ 2 \omega_{ \bd{q} } }
\left( 
b_{ -\bd{q} } + b^\dagger_{ \vphantom{-} \bd{q} }
\right)  
\left( 
b_{ \bd{q} } + b^\dagger_{ - \bd{q} }
\right) ,
\label{eq:HDebye2}
\end{align}
where we have introduced the bare electron-phonon coupling
\begin{equation}
g_{ \bd{q} } = 
\frac{ z \sqrt{ n_i } q 
}{ \sqrt{ 2 M \omega_{ \bd{q} } } } 
f_{ \bd{q} } 
= \frac{ \lambda_{ \bd{q} } 
}{ \sqrt{ 2 \omega_{ \bd{q} } } } 
f_{ \bd{q} } ,
\label{eq:gq}
\end{equation}
and the square of the ionic plasma frequency
\begin{equation}
\Omega_i^2 = \frac{ 4 \pi ( ze )^2 n_i }{ M } = \lambda_{\bd{q}}^2 f_{\bd{q}}.
\label{eq:Omegaidef}
\end{equation}
The last term in Eq.~\eqref{eq:HDebye2} represents the direct Coulomb interaction between
ionic charge fluctuations which
can be expressed in terms of the  ionic displacement operator 
$ X_{ \bd{q} } $ defined in Eq.~\eqref{eq:Xdef} as follows:
\begin{equation}
\mathcal{U}_{ i i } 
= \frac{ 1 }{ 2 } \sum_{ \bd{q} \neq 0 }
\Omega_i^2 X_{ - \bd{q} } X_{ \bd{q} } .
\label{eq:Uii} 
\end{equation}
In the textbook by Pines and Nozi\`{e}res \cite{Pines89} 
this contribution is simply ignored, while
Fetter and Walecka \cite{Fetter71} take it into account implicitly by replacing the long-range Coulomb interaction (by hand) 
by an effective short-range interaction.
However, as anticipated in Refs.~\cite{Kopietz96b,Kopietz97}
and recently pointed out again in Ref.~\cite{Hansen23},
for a microscopic derivation of the effective screened interaction
between electrons a more careful treatment of the ionic Coulomb interaction ${\cal{U}}_{ii}$ is necessary.
It is therefore not surprising that ${\cal{U}}_{ii}$ is also essential for obtaining the dispersions of the hybrid plasmon-sound modes of the Debye model.

\section{Exact equation for the hybrid plasmon-sound modes}

\label{sec:exact}

In the Debye model the displacement of the ions (phonons) is accompanied by a redistribution of the charge, 
so that phonons and plasmons are not independent.
The energy dispersion and the damping of the resultant collective excitations can be obtained from the poles of the phonon propagator in the
complex frequency plane. 
As outlined in the Appendix, 
the phonon propagator of the Debye model can be written in the following form:
\begin{equation}
D ( \bd{q} ,  \omega ) = 
\frac{ 1 }{ 
- \omega^2 + \omega_{ \bd{q} }^2 
+ W ( \bd{q} , \omega ) 
} .
\label{eq:DW}
\end{equation}
Here, 
the phonon self-energy 
$ W ( \bd{q} , {\omega} ) $ 
can be expressed in terms of the electronic contribution
$ \Pi_{ e e } ( \bd{q} , \omega ) $ 
to the interaction irreducible density  correlation function defined in Eq.~\eqref{eq:Pidef} below,
\begin{equation}
W ( \bd{q} , {\omega} ) 
= \frac{ \Omega_i^2 }{ 
1 + f_{ \bd{q} } \Pi_{ee} ( \bd{q} , {\omega}  ) 
} .
\label{eq:phonselfDebye}
\end{equation}
From the poles of the phonon propagator \eqref{eq:DW} in the complex frequency plane,
we obtain the following self-consistency equation for the renormalized phonon energies:
\begin{equation}
\omega^2 = 
\omega_{ \bd{q} }^2 + 
\frac{ \Omega_i^2 }{ 1 + 
f_{ \bd{q} } \Pi_{ e e } ( \bd{q} , \omega  ) } .
\label{eq:phononDebye}
\end{equation}
In the adiabatic limit where the phonons are slow compared to the electrons,  
the electronic polarization can be approximated by the electronic compressibility 
$ \partial n_e / \partial \mu $, 
i.e.,
\begin{equation}
\Pi_{ e e } ( \bd{q} , \omega ) \approx 
\Pi_{ e e } ( \bd{q} , 0 ) \approx 
\frac{ \partial n_e }{ \partial \mu } ,
\end{equation}  
where $ \mu $ is the chemical potential. 
Then Eq.~\eqref{eq:phononDebye}  reduces to the following expression
for the renormalized phonon energies:
\begin{equation}
\omega^2 \approx 
\omega_{ \bd{q} }^2 + 
\frac{ \Omega_i^2 }{ 1 + \kappa^2 / q^2 } ,
\end{equation}
where 
$ \kappa^2 = 4 \pi e^2 
\partial n_e / \partial \mu $ 
is the square of the Thomas-Fermi screening wavevector. 
For $ q \ll \kappa $ we thus obtain
\begin{equation}
\omega^2 \approx 
\omega_{ \bd{q} }^2 + 
\frac{ \Omega_i^2 }{ \kappa^2 } q^2 
= \left( 
c^2 + \frac{ z m }{ 3 M } v_F^2 
\right) q^2 ,
\end{equation}
with the Fermi velocity $ v_F $.
If the bare phonon velocity $ c $ is small compared with $ \Omega_i / \kappa $ 
we obtain the well-known Bohm-Staver 
relation \cite{Bohm51,Ashcroft76} for the
renormalized phonon velocity of a metal,
\begin{equation}
v = \sqrt{ \frac{ z m}{ 3 M } } v_F .
\end{equation}

It is instructive to compare  the exact equation \eqref{eq:phononDebye} 
with the perturbative formula for the renormalized phonon energy given in the book by Pines and Nozi\`{e}res \cite{Pines89}. 
To that end, 
we decompose the second term on the right-hand side of Eq.~\eqref{eq:phononDebye} as follows:
\begin{align}
&
\frac{ \Omega_i^2 
}{ 1 + 
f_{ \bd{q} } \Pi_{ e e } ( \bd{q} , \omega  ) 
}
\nonumber\\
= {} & 
\frac{ \Omega_i^2 
[ 1 +
f_{ \bd{q} } \Pi_{ e e } ( \bd{q} , \omega  ) -
f_{ \bd{q} } \Pi_{ e e } ( \bd{q} , \omega  )  
]   
}{ 1 + 
f_{ \bd{q} } \Pi_{ e e } ( \bd{q} , \omega  )  
}
\nonumber\\
= {} & 
\Omega_i^2  - 
\frac{ \lambda_{\bd{q}}^2 
f^2_{ \bd{q} } \Pi_{ e e } ( \bd{q} , \omega) 
}{ 1 + 
f_{ \bd{q} } \Pi_{ e e } ( \bd{q} , \omega  ) 
} .
\label{eq:omegadecomp}
\end{align}
If we simply omit the $ \Omega_i^2 $-term and approximate 
$ \Pi_{ e e } ( \bd{q} , \omega ) $
in the second term by its non-interacting limit 
$ \Pi_{0} ( \bd{q} , \omega ) $,
we obtain instead of Eq.~\eqref{eq:phononDebye}
\begin{equation}
\omega^2 \approx
\omega_{ \bd{q} }^2  - 
\frac{ 
\lambda_{ \bd{q} }^2 
f^2_{ \bd{q} } \Pi_{0} ( \bd{q} , \omega ) 
}{ 1 + 
f_{ \bd{q} } \Pi_{0} ( \bd{q} , \omega  ) } ,
\label{eq:phononPines}
\end{equation}
which is equivalent to the perturbative expression for the
renormalized phonon frequencies given 
in Eq.~(4.100) of Ref.~\cite{Pines89}.
Obviously,  
in Eq.~\eqref{eq:phononPines}
the first contribution involving
$ \lambda_{ \bd{q} }^2 f_{ \bd{q} } 
= \Omega_i^2 $ on the right-hand side of Eq.~\eqref{eq:omegadecomp} is missing.
The reason for this discrepancy is that the Hamiltonian considered in  Ref.~\cite{Pines89} 
does not include the direct Coulomb interaction 
$ \mathcal{U}_{ i i } $ 
between the ionic charge fluctuations represented by the last term in Eq.~\eqref{eq:HDebye2}. 
At first glance,
it seems reasonable to neglect this term 
for the calculation of the hybrid plasmon-sound modes
because the ionic plasma frequency
$\Omega_i$ is usually small compared with the 
 electronic plasma frequency
$ \Omega_e $ defined by
 \begin{equation}
 \Omega_e^2 = \frac{ 4 \pi e^2 n_e}{m}.
 \label{eq:Omegaedef}
 \end{equation}
However, for the calculation of the hybridized collective charge modes
the direct Coulomb interaction between the ionic charge fluctuations represented by last term in Eq.~\eqref{eq:HDebye2} cannot be neglected because without this term 
the solution of the resulting  equation~\eqref{eq:phononPines} for the mode energies does not have a sound-like solution  with linear dispersion for small momentum, as illustrated in Fig.~\ref{fig:phonondispersions} below.

Given the fact that phonons in the Debye model are accompanied by  charge fluctuations, 
it is also possible to obtain our equation \eqref{eq:phononDebye} for the hybrid plasmon-phonon modes from the poles of the charge-correlation functions. 
As shown in the Appendix,
the  correlation function 
$ \chi ( \bd{q} , \omega ) $ 
of the total charge density of the Debye model can be written as
\begin{equation}
\chi ( \bd{q} , \omega ) = 
\frac{ \Pi ( \bd{q} , \omega ) 
}{ 1 + f_{ \bd{q} } \Pi ( \bd{q} , \omega ) } ,
\label{eq:chidef}
\end{equation}
where the total irreducible polarization  
$ \Pi ( \bd{q} , \omega ) $ 
is the sum of an electronic and an ionic contribution,
\begin{equation}
\Pi ( \bd{q} , \omega ) = 
\Pi_{ e e } ( \bd{q} , \omega ) + 
\Pi_{ i i } ( \bd{q} , \omega ) .
\label{eq:Pidef}
\end{equation}
While the electronic contribution 
$ \Pi_{ e e } ( \bd{q} , \omega ) $ 
cannot be calculated exactly and 
is only known in certain limits 
(for example, 
in the hydrodynamic regime, 
see Sec.~\ref{sec:hydro}), 
we show in the Appendix
that the ionic contribution is simply given by
\begin{equation}
\Pi_{ i i } ( \bd{q} , \omega ) =
\frac{ \lambda_{ \bd{q} }^2 
}{ 
\omega_{ \bd{q} }^2 - \omega^2 
} 
= \frac{ z^2 n_i q^2 / M 
}{ 
\omega_{ \bd{q} }^2 - \omega^2 
} .
\label{eq:Piii}
\end{equation}
Since the total charge density can be decomposed into an electronic and an ionic part, 
$ \rho = \rho_e - \rho_i $, 
the total charge correlation function can be written as
\begin{align}
\chi ( \bd{q} , \omega ) 
= {} &
\chi_{ e e } ( \bd{q} , \omega ) + 
\chi_{ i i } ( \bd{q} , \omega ) 
\nonumber\\
&
- 
\chi_{ e i } ( \bd{q} , \omega ) - 
\chi_{ i e } ( \bd{q} , \omega ) ,
\label{eq:chitot}
\end{align}
with
\begin{subequations}
\begin{align}
\chi_{ e e } ( \bd{q} , \omega ) 
& = 
\frac{ 
\Pi_{ e e } ( \bd{q} , \omega  ) 
[ 1 + 
f_{ \bd{q} } \Pi_{ i i } ( \bd{q} , \omega  ) 
] }{ 
1 + f_{ \bd{q} } [ 
\Pi_{ e e } ( \bd{q} , \omega ) + 
\Pi_{ i i } ( \bd{q} , \omega ) 
] } ,
\label{eq:chiee}
\\
\chi_{ i i } ( \bd{q} , \omega  ) 
& = \frac{ 
\Pi_{ i i } ( \bd{q} , \omega  ) 
[ 1 + 
f_{ \bd{q} } \Pi_{ e e } ( \bd{q} , \omega ) 
] }{ 
1 + f_{ \bd{q} } [ 
\Pi_{ e e } ( \bd{q} , \omega ) + 
\Pi_{ i i } ( \bd{q} , \omega ) 
] } ,
\label{eq:chiii}
\\
\chi_{ e i } ( \bd{q} , \omega  ) 
& =  
\frac{ 
f_{ \bd{q} } 
\Pi_{ e e } ( \bd{q} , \omega ) 
\Pi_{ i i } ( \bd{q} , \omega )
}{ 
1 + f_{ \bd{q} } [ 
\Pi_{ e e } ( \bd{q} , \omega ) + 
\Pi_{ i i } ( \bd{q} , \omega ) 
] }  
\nonumber\\
& = \chi_{ i e } ( \bd{q} , \omega) .
\label{eq:chiei}
\end{align}
\end{subequations}
The collective charge modes of the system can be obtained from the poles of the charge correlation functions in the complex frequency plane, 
i.e.,
\begin{equation}
1 +  f_{\bd{q}} [ 
\Pi_{ e e } ( \bd{q} , \omega ) + 
\Pi_{ i i } ( \bd{q} , \omega ) ] 
= 0 .
\end{equation}
Because the ionic polarization is given exactly by Eq.~\eqref{eq:Piii},
this self-consistency equation is equivalent to Eq.~\eqref{eq:phononDebye}
obtained from the phonon propagator \eqref{eq:DW}.

\section{Plasmon-sound modes in the hydrodynamic regime}

\label{sec:hydro}

\begin{figure}[tb]
\includegraphics[width=\linewidth]{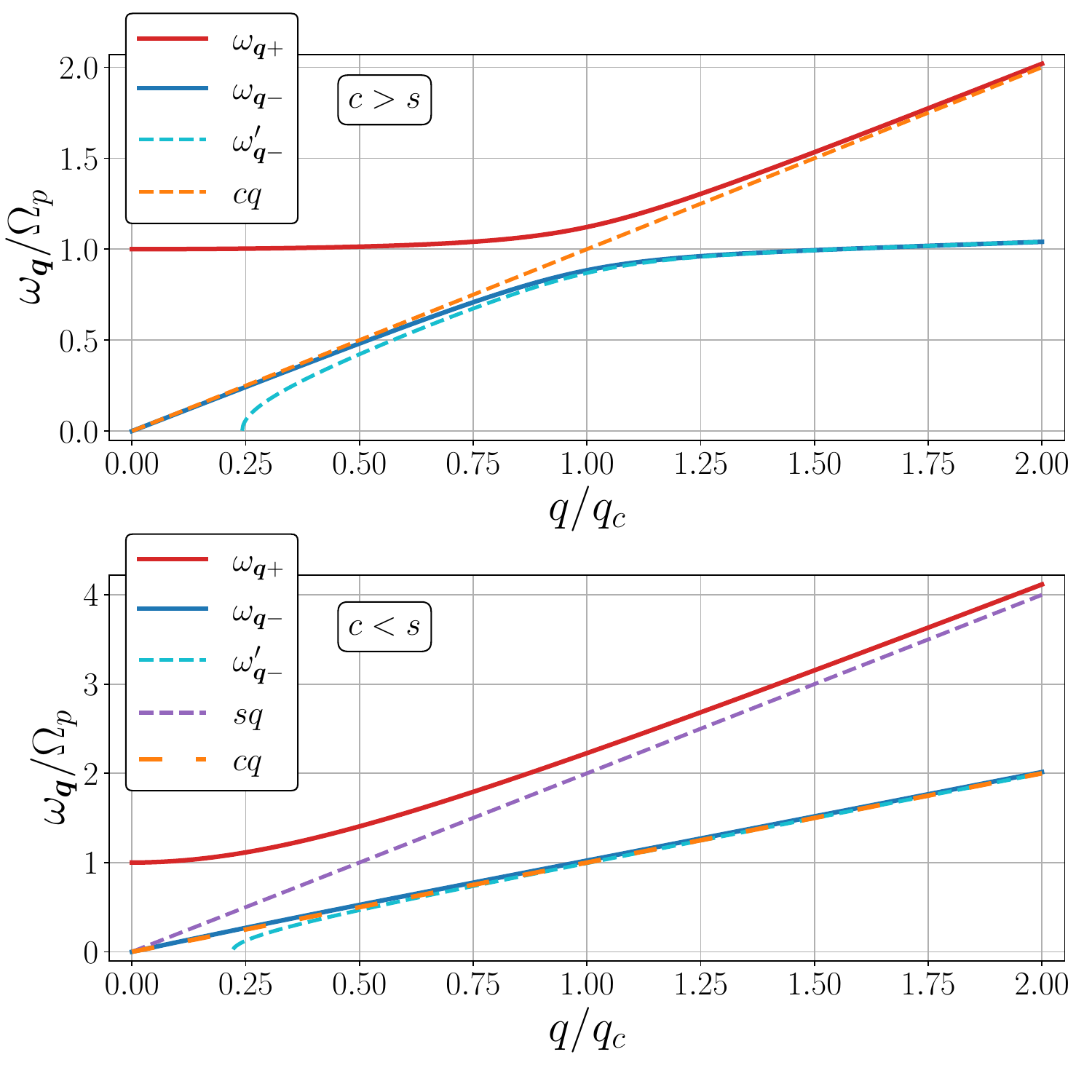}
\caption{The plasmon-sound dispersions 
$ \omega_{ \bd{q} \pm } $ 
given in Eq.~\eqref{eq:omegapms} 
as a function of the wavevector $ q $ for 
$ c > s $ (upper panel) and 
$ c < s $ (lower panel).
For comparison,
we also show the bare phonon and electronic sound dispersions,
as well as the low-energy solution 
$ \omega_{ \bd{q} - }' $
of the approximate self-consistency equation \eqref{eq:phononPines}
that does not take the direct ionic Coulomb interaction 
$ \mathcal{ U }_{ i i } $
into account.
The latter exhibits an unphysical long-wavelength behavior: 
Instead of approaching the bare phonon dispersion asymptotically for $ q \to 0 $ like the complete solution
$ \omega_{ \bd{q} - } $,
it vainishes  at a finite wavevector $ q > 0 $.
The plots are for 
$ \Omega_e = 4 \Omega_i $,
and
$ c / s = 5 $ and $ 1 / 2 $
for the upper and lower panels,
respectively.
The crossover wavevector where there is strong plasmon-phonon mixing for $ c > s $
is denoted by $ q_c = \Omega_p / c $ where $\Omega_p = \sqrt{\Omega_i^2 + \Omega_e^2}$ is the total plasma frequency
}
\label{fig:phonondispersions}
\end{figure}

\begin{figure}[tb]
\includegraphics[width=\linewidth]{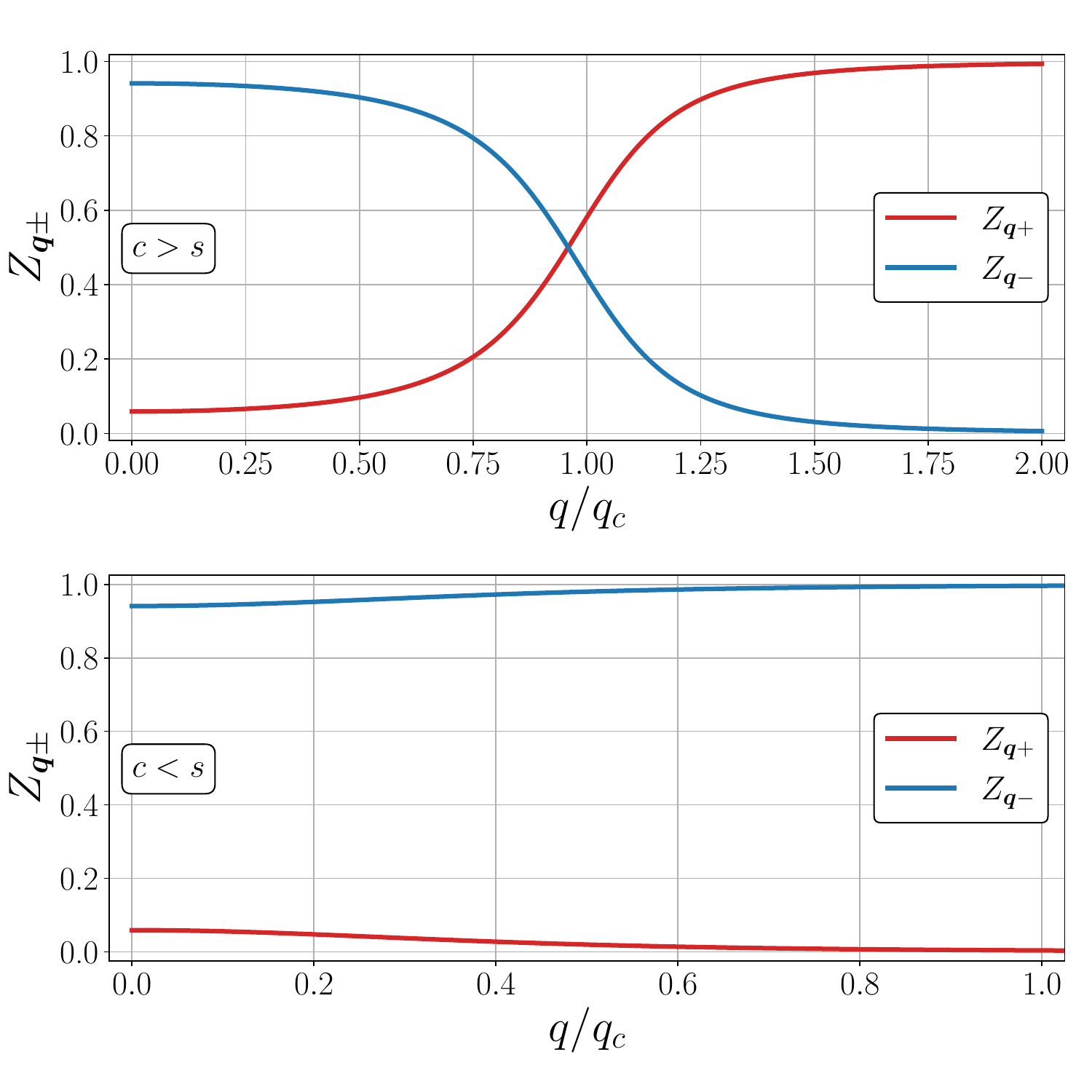}
\caption{Graph of the dimensionless residues  
$ Z_{ \bd{q} \pm } $ given in Eqs.~\eqref{eq:Dzz}
of the two modes 
$ \omega_{ \bd{q} \pm } $ 
in the phonon propagator \eqref{eq:DW},
for 
$ c > s $ (upper panel) and
$ c < s $ (lower panel). 
Note that by construction 
$ Z_{ \bd{q} + } + Z_{ \bd{q} - } = 1 $. 
Parameters are the same as in Fig.~\ref{fig:phonondispersions}
}
\label{fig:residues}
\end{figure}

\begin{figure}[tb]
 \includegraphics[width=\linewidth]{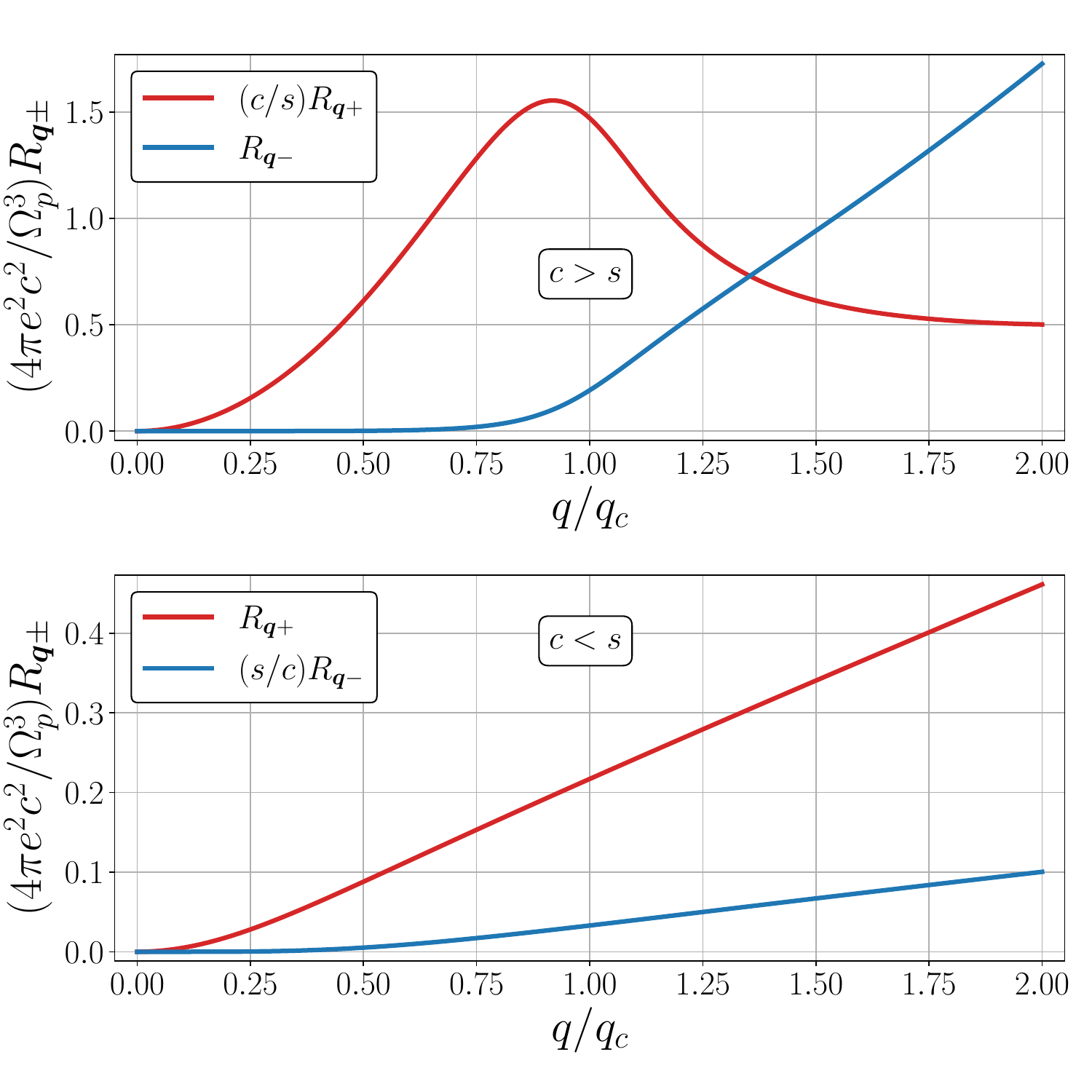}
\caption{Graph of the weights $ R_{ \bd{q} \pm } $
of the two modes $ \omega_{ \bd{q} \pm } $ in the dynamic structure factor
$ S ( \bd{q} , \omega ) $ 
in the hydrodynamic regime 
for
$ c > s $ (upper panel) and
$ c < s $ (lower panel), see Eq.~\eqref{eq:Sres}.
Note that for large wavevectors,
both weights increase linearly with $ q $.
Parameters are the same as in Fig.~\ref{fig:phonondispersions}
}
\label{fig:weights}
\end{figure}

For the explicit calculation of the hybrid modes, 
we need the electronic polarization
$ \Pi_{ e e } ( \bd{q} , \omega ) $  
of the Debye model, 
which in general cannot be calculated exactly.
In this work we focus on the hydrodynamic regime,
where rapid collisions generate a state of local equilibrium.
Then 
$ \Pi_{ e e } ( \bd{q} , \omega ) $  
has the following general form for small momenta and frequencies \cite{Pines89}:
\begin{equation}
\Pi_{ e e } ( \bd{q} , \omega ) = 
\frac{ n_e q^2 / m }{ 
\nu_{ \bd{q} }^2 - \omega^2 } ,
\label{eq:Piee}
\end{equation}
where
\begin{equation}
\nu_{ \bd{q} } = s q 
\end{equation}
is the frequency of electronic sound with velocity $ s $. 
Actually, 
as pointed out by Pines and  Nozi\`{e}res \cite{Pines89},
the expression \eqref{eq:Piee} is also valid in the strong-coupling collisionless regime 
where the collisionless (Landau) damping of the electronic sound mode can be neglected.  
Using the fact that Eq.~\eqref{eq:Piee} implies
\begin{equation}
f_{ \bd{q} } \Pi_{ e e } ( \bd{q} , \omega )  
= 
\frac{ \Omega_e^2 }{ 
\nu_{ \bd{q} }^2  - \omega^2 } ,
\end{equation}
where the square of the electronic plasma frequency 
$\Omega_e^2$
is defined in Eq.~(\ref{eq:Omegaedef}),
%
%
%\begin{equation}
%
%
%\Omega_e^2 = \frac{4 \pi e^2 n_e}{m} ,
%
%
%\label{eq:Omegaedef}
%
%
%\end{equation}
%
%
we conclude that
our self-consistency equation \eqref{eq:phononDebye} for the collective charge modes of the Debye model reduces to
\begin{equation}
\omega^2   = 
\omega_{ \bd{q} }^2 + 
\frac{ 
\Omega_i^2 
}{ 
1 + \frac{ 
\Omega_e^2 
}{ 
\nu_{ \bd{q} }^2 -  \omega^2 
} 
} .
\label{eq:selfcon2}
\end{equation}
The solutions of this bi-quadratic equation are
\begin{align}
\omega_{ \bd{q}  \pm }^2  
= {} &    
\frac{ 1 }{ 2 } \left( 
\omega_{ \bd{q} }^2 + 
\nu_{ \bd{q} }^2 + 
\Omega_i^2 + 
\Omega_e^2 
\right)
\nonumber\\
\pm   &
\sqrt{
\frac{ 1 }{ 4 } 
\left( 
\omega_{ \bd{q} }^2 - 
\nu_{ \bd{q} }^2 + 
\Omega_i^2 - 
\Omega_e^2 
\right)^2
+ \Omega_i^2 \Omega_e^2 
} .
\label{eq:omegapms}
\end{align}
Note that the equation for the energies of the collective modes arising from the hybridization between phonons and phase fluctuations in a quasi-one dimensional electronic charge-density wave derived in Ref.~\cite{Hansen23} has exactly the same structure.
A graph of 
$ \omega_{ \bd{q} + } $ and 
$ \omega_{ \bd{q} - } $
 as a function of the momentum $ q $ is shown in Fig.~\ref{fig:phonondispersions}.
The asymptotic behavior of 
$ \omega_{ \bd{q} \pm }^2 $ 
for small $ q $ is
\begin{subequations}
\begin{align}
\omega_{ \bd{q} + }^2 
& = 
\Omega_p^2 + 
\frac{ \Omega_i^2 }{ \Omega_p^2 } 
\omega_{ \bd{q} }^2 +
\frac{ \Omega_e^2 }{ \Omega_p^2 } 
\nu_{\bd{q}}^2 + \mathcal{O} ( q^4 ) ,
\\
\omega_{ \bd{q} - }^2 
& = 
\frac{ \Omega_e^2 }{ \Omega_p^2 } 
\omega_{ \bd{q} }^2 +
\frac{ \Omega_i^2 }{ \Omega_p^2 } 
\nu_{ \bd{q} }^2 + \mathcal{O} ( q^4 ) ,
\end{align}
\end{subequations}
where
\begin{equation}
\Omega_p^2 = \Omega_e^2 + \Omega_i^2 
\end{equation}
is the square of the total plasma frequency.
Obviously, 
long-range Coulomb interactions do not prohibit the existence of a sound mode  
$ \omega_{ \bd{q} - } $ with linear dispersion; 
the sound velocity
\begin{equation}
v = \sqrt{ 
\frac{ 
\Omega_e^2 c^2 + \Omega_i^2 s^2
}{
\Omega_e^2 + \Omega_i^2 
} 
}
\end{equation}
is a weighted geometric average of the phononic and electronic sound velocities, 
with weights determined by the ratio of the electronic and the ionic plasma frequencies.
Under the realistic assumption
$ \Omega_i^2 \ll \Omega_e^2 $
where $ v \approx c $,
an avoided level crossing of plasmon and phonon modes and hence strong hybridization is only possible for $ c > s $ and occurs for wavevectors 
$ q \approx q_c = \Omega_p / c $.
In the opposite case $ c < s $ there is no avoided level crossing and only weak mixing between
plasmons and phonons.
These two generic scenarios are shown in the upper and lower panels of Fig.~\ref{fig:phonondispersions},
respectively.
On the other hand, 
for large momenta when both 
$ \omega_{ \bd{q} } $ and 
$ \nu_{ \bd{q} } $
are large compared with the plasma frequencies
we obtain to next-to-leading order
\begin{subequations}
\begin{align}
\omega_{\bd{q} + }^2 
& = 
\textrm{max} \{ 
\omega_{ \bd{q} }^2 + \Omega_i^2 , 
\nu_{ \bd{q} }^2 + \Omega_e^2 \} ,
\\
\omega_{ \bd{q} - }^2 
& = \textrm{min} \{ 
\omega_{ \bd{q} }^2 + \Omega_i^2 , 
\nu_{ \bd{q} }^2 + \Omega_e^2 \} .  
\end{align}
\end{subequations}
Given the collective modes \eqref{eq:omegapms},
we can write the phonon propagator \eqref{eq:DW} in the hydrodynamic regime in the form
\begin{align}
D( \bd{q} , \omega ) 
& = 
\frac{ 
\Omega_e^2 + \nu_{ \bd{q} }^2 - \omega^2 
}{ 
\left( \omega_{ \bd{q} + }^2 - \omega^2 \right)
\left( \omega_{ \bd{q} - }^2 - \omega^2 \right)
}
\nonumber\\
& = 
\frac{ Z_{ \bd{q} + } 
}{ \omega_{ \bd{q} + }^2 - \omega^2 } +
\frac{ Z_{ \bd{q} - } 
}{ \omega_{ \bd{q} - }^2 - \omega^2 } ,
\end{align}
with the dimensionless residues
\begin{subequations}
\label{eq:Dzz}
\begin{align}
Z_{ \bd{q} + } & = 
\frac{ 
\omega_{ \bd{q} + }^2 
- \Omega_e^2 
- \nu_{ \bd{q} }^2 
}{
\omega_{ \bd{q} + }^2 - 
\omega_{ \bd{q} - }^2 
} ,
\\
Z_{ \bd{q} - } 
& = 
\frac{ 
\Omega_e^2 
+ \nu_{ \bd{q} }^2 
- \omega_{ \bd{q} - }^2 
}{
\omega_{ \bd{q} + }^2 - 
\omega_{ \bd{q} - }^2 } .
\end{align}
\end{subequations}
The momentum dependence of these residues is
shown graphically in Fig.~\ref{fig:residues}.
For $ c > s $,
they display strong hybridization in the crossover regime $ q \approx q_c $,
where the lower hybrid mode switches from being phonon-like to plasmon-like.
On the other hand,
for $ c < s $ there is only weak mixing and almost all of the spectral weight in the phonon propagator remains in the lower,
phonon-like hybrid mode.
By combining Eqs.~(\ref{eq:chidef}--\ref{eq:Piii}) with the hydrodynamic expression  for the electronic polarization 
$ \Pi_{ e e } ( \bd{q} , \omega ) $
given in  Eq.~\eqref{eq:Piee},
we can also calculate the dynamic structure factor, 
i.e., 
the spectral function of charge fluctuations.
For $ \omega > 0 $ 
we obtain at finite  temperature $ T $,
\begin{align}
S ( \bd{q} , \omega ) 
= {} &   
\frac{ 1 }{ 1 - e^{ - \omega / T } }
\frac{ 1 }{ \pi } \textrm{Im} \, 
\chi ( \bd{q} , \omega + i 0^+ ) 
\nonumber\\
= {} & 
\frac{ 1 }{ 1 - e^{ - \omega / T } } 
\bigl[ 
R_{ \bd{q} + }
\delta \left( 
\omega - \omega_{ \bd{q} + } 
\right)
\nonumber\\
& \hspace{1.7cm} 
+ R_{ \bd{q} - }
\delta \left( 
\omega - \omega_{ \bd{q} - } 
\right) \bigr] ,
\label{eq:Sres}
\end{align}
where the weights are  
\begin{equation}
R_{ \bd{q} \pm } = 
\frac{ 
\left( \Omega_e^2  + \Omega_i^2 \right)
\omega_{ \bd{q} \pm }^2 -
\Omega_e^2 \omega_{ \bd{q} }^2 - 
\Omega_i^2 \nu_{ \bd{q} }^2 
}{
2 f_{ \bd{q} } \omega_{ \bd{q} \pm }
\left( 
\omega_{ \bd{q} \pm }^2 - 
\omega_{ \bd{q} \mp }^2 
\right) } .
\end{equation}
These weights are shown in Fig.~\ref{fig:weights}.
While  for $ q \gtrsim q_c $ both weights  grow linearly with
$q$, for $ q < q_c $ the phonon-like lower hybrid mode gives only a vanishingly small contribution to the dynamic structure factor.
For $ c < s $ where the modes hybridize only weakly
the weight of the lower mode remains small also for $ q > q_c$.
In contrast,
for $ c > s $ one clearly observes a transfer of spectral weight from the upper to the lower hybrid mode close to the crossover wavevector
$ q \approx q_c $.
For negative frequencies the dynamic structure factor can be obtained from detailed balance,
$ S ( \bd{q} , - \omega ) 
= e^{ -  \omega / T  } 
S ( \bd{q} , \omega ) $. 

\section{Hydrodynamic damping of plasmon-sound modes}

\begin{figure}[tb]
\includegraphics[width=\linewidth]{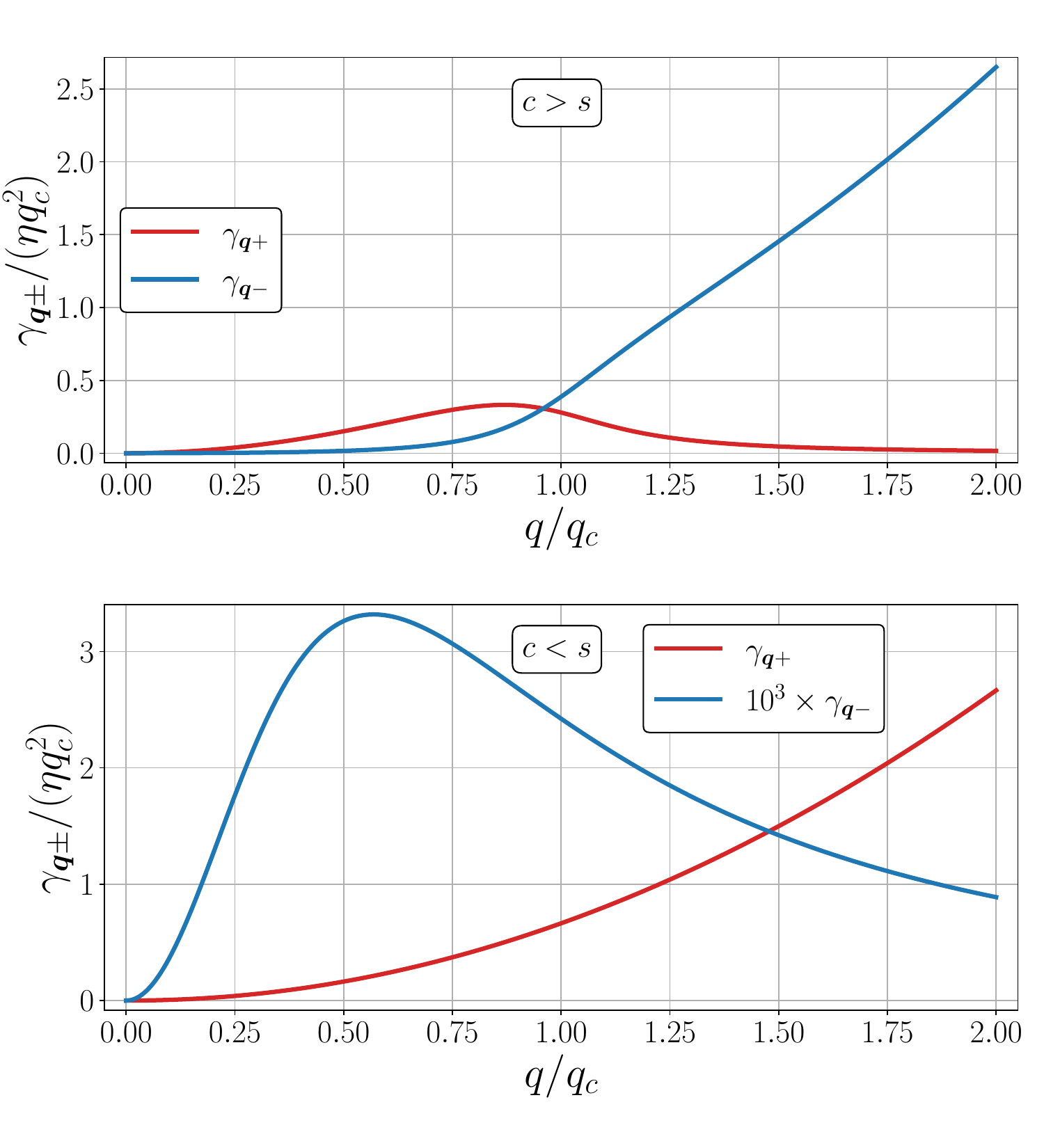}
\caption{Damping rates 
$ \gamma_{ \bd{q} \pm } $ 
of the hybrid plasmon-sound modes in the hydrodynamic regime,
Eqs.~\eqref{eq:damping},
as a function of wavevector $ q $
for $ c > s $ (upper panel)
and $ c < s $ (lower panel).
Note that the phonon-like mode remains rather sharp,
whereas the plasmon-like mode disappears for large $ q $.
The parameters are the same as in Fig.~\ref{fig:phonondispersions}}
\label{fig:damping}
\end{figure}

The hydrodynamic form of 
$ \Pi_{ e e } ( \bd{q} , \omega ) $ in Eq.~\eqref{eq:Piee} does not contain dissipative terms, 
which are of higher order in 
$ \omega $ and $ q $. 
To calculate the lifetime of the
collective modes we should include the leading dissipative term in the hydrodynamic regime, 
which leads to the following modification of Eq.~\eqref{eq:Piee} \cite{Pines89,Forster75},
\begin{equation}
\Pi_{ e e } ( \bd{q} , \omega ) =
\frac{ n q^2 / m }{  
\nu_{ \bd{q} }^2  - 
\omega^2 -  
i \omega  / \tau_{q} } .
\label{eq:Pihydro3}
\end{equation}
At long wavelengths we may furthermore expand the damping rate $ 1/ \tau_q $ up to quadratic order in $ q $,
\begin{equation}
\frac{ 1 }{ \tau_q } = 
\frac{ 1 }{ \tau_0 } + D q^2 .
\label{eq:tauq}
\end{equation}
The momentum-independent part  $ 1/ \tau_0 $ can  be expressed in terms of the static conductivity $ \sigma $ via the Drude formula
$ \sigma = n_e e^2 \tau_0 / m $, 
implying
\begin{equation}
\frac{1}{ \tau_0 } =  
\frac{ \Omega_e^2 }{ 4 \pi \sigma } .
\end{equation} 
The coefficient $ D $ of the $ q^2 $-term in the damping rate \eqref{eq:tauq} can be expressed in terms of the longitudinal ($ D_l $) and transverse ($ D_t $) diffusion coefficients as follows~\cite{Forster75}:
\begin{equation}
D 
= 
D_l + D_t 
\left( \frac{ c_p }{ c_V } - 1 \right) 
= 
\frac{4}{3} \eta + \zeta + D_t 
\left( \frac{ c_p }{ c_V } - 1 \right) ,
\end{equation}
where $ c_p $ and $ c_V $ are heat capacities at constant pressure and volume, 
and we have expressed the longitudinal diffusion coefficient $ D_l $ in terms of the (kinematic) shear viscosity $ \eta $ and bulk viscosity $ \zeta $. 
In a system where the contribution from the shear viscosity dominates, 
we obtain
\begin{equation}
\frac{ 1 }{ \tau_q } = 
\frac{ \Omega_e^2 }{  4 \pi \sigma } + 
\frac{ 4 }{ 3 } \eta q^2 .
\end{equation}
Because the Debye Hamiltonian \eqref{eq:HDebye} is translationally invariant
the total momentum is conserved, 
so that the static conductivity is infinite and hence  
$ 1 / \tau_0 = 0 $ and 
$ 1 / \tau_q = D q^2 $.
The electronic irreducible polarization \eqref{eq:Pihydro3} then reduces to
\begin{equation}
\Pi_{ e e } ( \bd{q} , \omega) =
\frac{ n q^2 / m }{  
\nu_{\bd{q}}^2 - 
\omega^2 -  
i D q^2 \omega } .
\label{eq:Pihydro4}
\end{equation}
Note that this implies that for finite frequencies the conductivity is
\begin{align}
\sigma ( \omega ) 
& = 
- i \omega e^2 
\lim_{ q \to 0 } 
\frac{ 
\Pi_{ e e } ( \bd{q} , \omega ) + 
\Pi_{ i i } ( \bd{q} , \omega ) 
}{ q^2 }
\nonumber\\
& = 
- \frac{ \Omega_e^2 + \Omega_i^2 
}{ 4 \pi i \omega } .
\end{align}
Substituting Eq.~\eqref{eq:Pihydro4} into our self-consistency equation \eqref{eq:phononDebye}
for the collective density modes of the Debye model we obtain instead of Eq.~\eqref{eq:selfcon2},
\begin{equation}
\omega^2  = 
\omega_{ \bd{q} }^2 + 
\frac{ \Omega_i^2 }{ 1 +  
\frac{ \Omega_e^2 }{ 
\nu_{\bd{q}}^2 -  
\omega^2 - 
i D q^2 \omega } } .
\label{eq:selfcon3}
\end{equation}
The solutions 
$ \tilde{ \omega }_{ \bd{q} \pm } $ 
of this equation now have a finite imaginary part. 
For small wavevectors $ q $ 
the damping rate $ 1 / \tau_q = D q^2 $ 
is small compared with the real part of the energies. 
Thus, 
we may calculate the solutions perturbatively in powers of $ 1/ \tau_q $. 
To first order, we obtain
\begin{equation}
\tilde{ \omega }_{ \bd{q} \pm } =
\omega_{\bd{q} \pm } - 
i \gamma_{ \bd{q} \pm } ,
\end{equation}
where $ \omega_{ \bd{q} \pm } $ 
are the real solutions for vanishing damping
given in Eq.~\eqref{eq:omegapms}, 
and the damping rates are
\begin{subequations}
\label{eq:damping}
\begin{align}
\gamma_{\bd{q} + } 
& = 
\frac{ 1 }{ 2 \tau_{q} } \, 
\frac{ \omega_{ \bd{q} + }^2 
- \Omega_i^2 
- \omega_{ \bd{q} }^2 
}{
\omega_{ \bd{q} + }^2 - 
\omega_{ \bd{q} - }^2 } ,
\\
\gamma_{ \bd{q} - } 
& = 
\frac{ 1 }{ 2 \tau_{q} } \, 
\frac{ \Omega_i^2 
+ \omega_{ \bd{q} }^2 
- \omega_{ \bd{q} - }^2 
}{
\omega_{ \bd{q} + }^2 - 
\omega_{ \bd{q} - }^2 } .
\end{align}
\end{subequations}
These damping rates are displayed in Fig.~\ref{fig:damping} as function of momentum.
For $ q \to 0 $, 
they reduce to
\begin{subequations}
\begin{align}
\gamma_{ \bd{q} + } 
& = 
\frac{ D q^2 }{ 2 } \,
\frac{ \Omega_e^2  }{
\Omega_e^2 + \Omega_i^2 } ,
\\
\gamma_{ \bd{q} - } 
& = 
\frac{ D q^2 }{ 2 } 
\frac{ \Omega_i^2 }{ 
\Omega_e^2 + \Omega_i^2 } .
\end{align}
\end{subequations}
Since in our model dissipation is introduced only 
in the electronic part  of the irreducible polarization,
the phonon-like mode remains rather sharp for  almost all wavevectors.
On the other hand,
the damping of the plasmon-like mode is dominated by the viscous damping 
$ \propto \eta q^2 $.
In consequence, 
the plasmon-like mode disappears at large $ q $.
Both modes have a comparable width only for 
$ c > s $ and  wavevectors in the regime 
$ q \approx q_c $, 
where the hybridization is strong.
The corresponding dynamic structure factor consists in this approximation of two displaced Lorentzians,
\begin{align}
S ( \bd{q} , \omega ) \approx 
\frac{ 1 }{ 1 - e^{ -  \omega / T } } \,
\frac{ 1 }{ \pi } 
&  
\left[  
\frac{ 
R_{ \bd{q} + } \gamma_{ \bd{q} + } 
}{ 
\left( \omega - \omega_{ \bd{q} + } \right)^2 
+ \gamma_{ \bd{q} + }^2 } 
\right.
\nonumber\\
& \left. +  
\frac{ 
R_{ \bd{q} - } \gamma_{ \bd{q} - } 
}{ 
\left( \omega - \omega_{ \bd{q} - } \right)^2 
+ \gamma_{ \bd{q} - }^2 }  
\right] .
\label{eq:Sreslor}
\end{align}
In the regime where this approximation is not accurate, 
the lineshape of the dynamic structure factor can be obtained from
\begin{align}
S ( \bd{q} , \omega ) = {} &
\frac{ 1 }{ 1 - e^{ -  \omega /T } } \, 
\frac{ 1 }{ \pi f_{ \bd{q} } } 
\nonumber\\
& \times
\textrm{Im} \left[ 
\frac{ 
\frac{ \Omega_e^2 }{ 
\nu_{ \bd{q} }^2 - 
\omega^2 - 
i D q^2 \omega }
+ \frac{ \Omega_i^2 }{ 
\omega_{ \bd{q} }^2 - 
\omega^2 } 
}{
1 +     
\frac{ \Omega_e^2 }{ 
\nu_{\bd{q}}^2 - 
\omega^2 - 
i D q^2 \omega }
+ \frac{ \Omega_i^2 }{ 
\omega_{ \bd{q} }^2 - \omega^2 } 
}
\right] .
\label{eq:dynstrucres}
\end{align}
\begin{figure*}[tb]
\includegraphics[width=.32\linewidth]{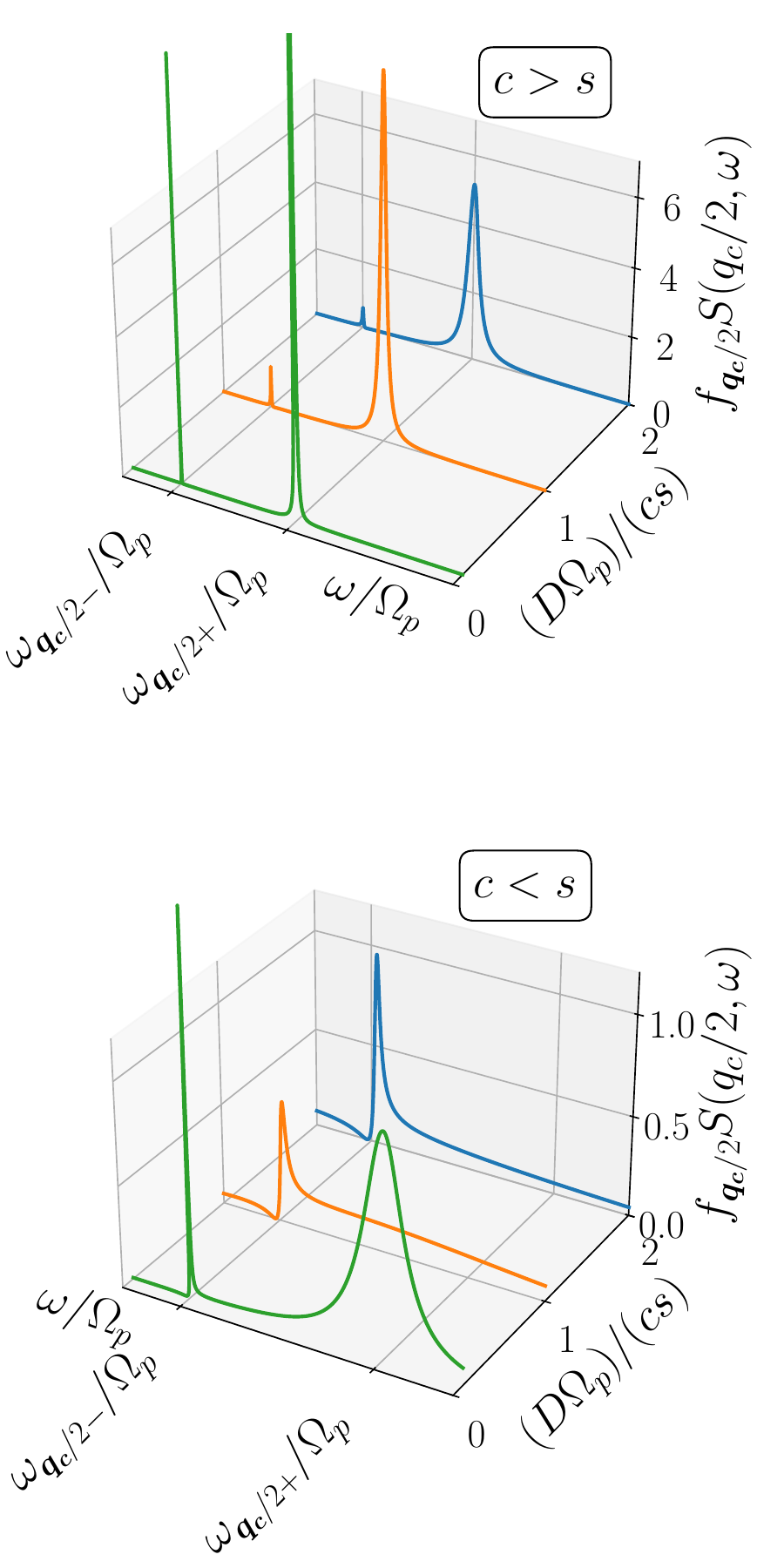}
\includegraphics[width=.32\linewidth]{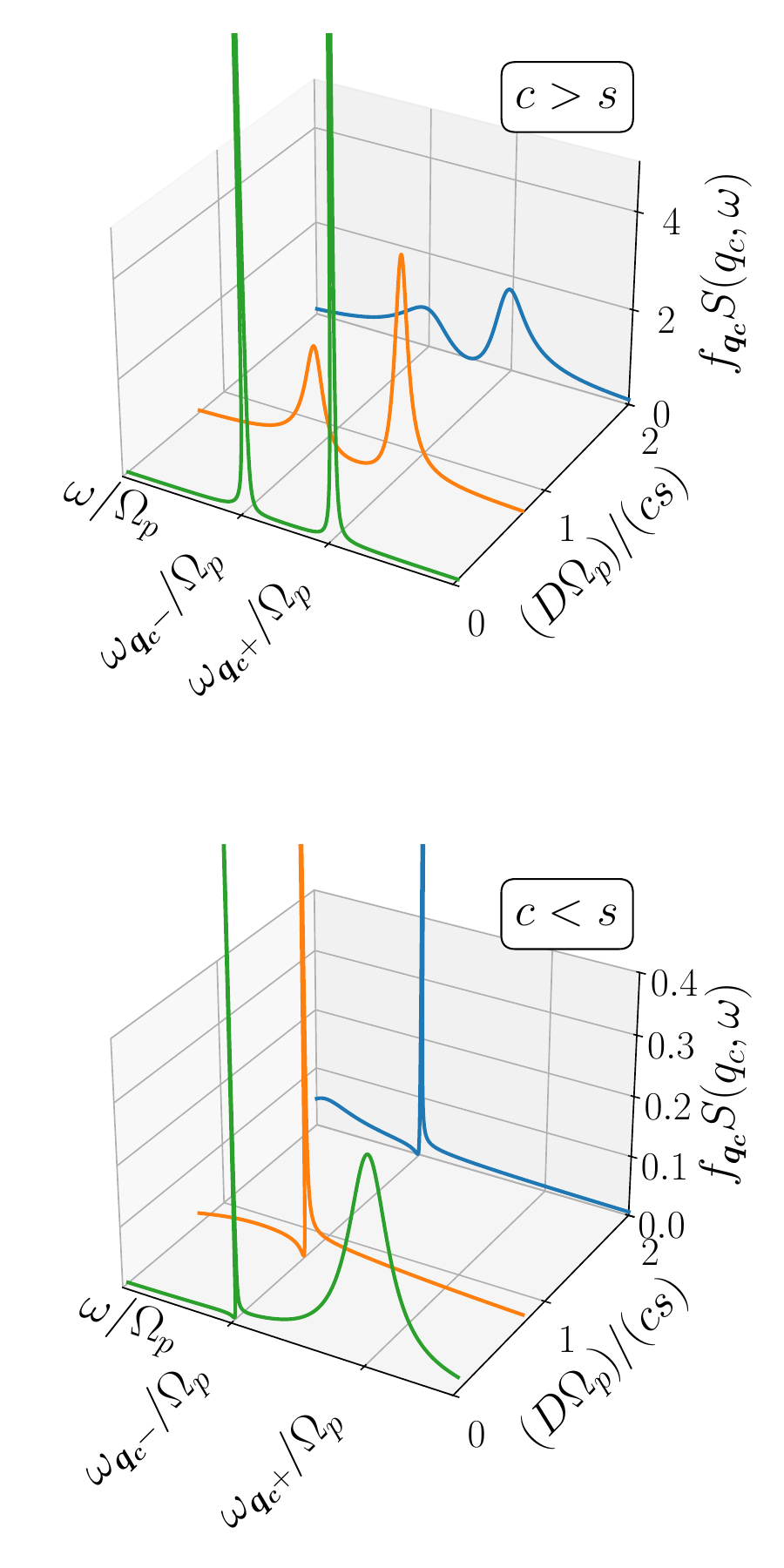}
\includegraphics[width=.32\linewidth]{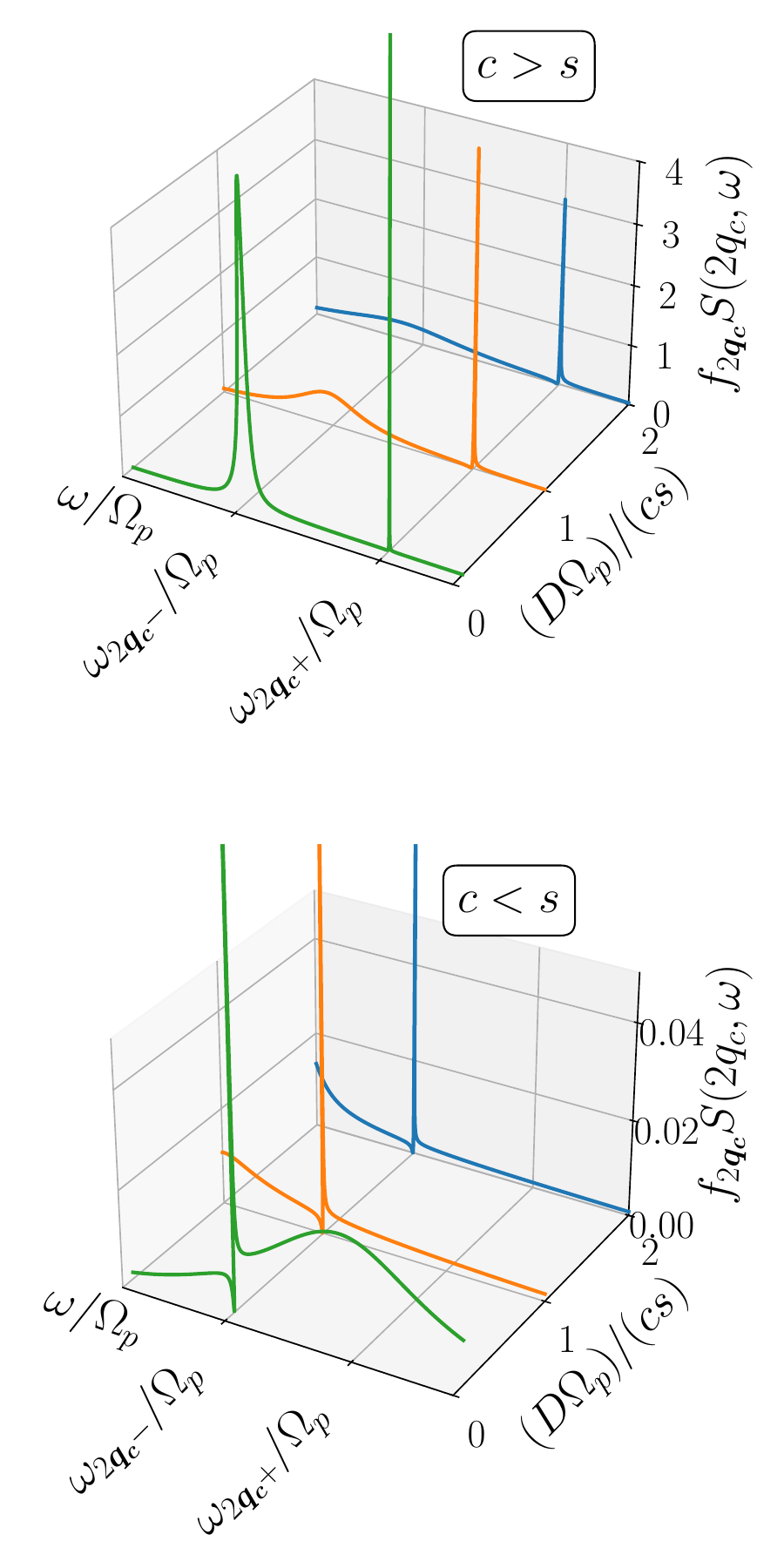}
\caption{Evolution of the spectral lineshape of the dynamic structure factor 
$ S ( \bd{q} , \omega ) $ 
given in Eq.~\eqref{eq:dynstrucres},
for $ c > s $ (upper panels) 
and $ c < s $ (lower panels),
and for different values of $q$,
with $ q = q_c / 2 $, $ q_c $, and $ 2 q_c $
(left, center, right).
The other parameters are the same as in Fig.~\ref{fig:phonondispersions}.
}
\label{fig:dynstruc}
\end{figure*}
Representative plots of the evolution of the spectral lineshape are displayed in Fig.~\ref{fig:dynstruc}.
From these plots we conclude that the plasmon-like mode quickly decays into a broad, incoherent background with increasing viscosity,
while the phonon-like mode remains rather sharp throughout.
However,
even in the weakly-interacting $ c < s $ case,
the overlap of the phonon peak with the incoherent remnant of the plasmon mode leads to an overall broadening and decrease of the phonon spectral weight,
which is particularly pronounced at long wavelengths.
%Only in the regime of strong hybridization
%at $ c > s $ and $ q \approx q_c $ 
%both hybrid modes have a significant  width.       

%In a system where the total momentum is not conserved (for example due to impurities or Umkapp %scettring) the damping rate $1/ \tau_q$ in Eq.~(\ref{eq:Pihydro3})
%has the more general form $1/ \tau_q = 1/ \tau_0 + \Gamma q^2 $.
%The momentum-independent part  $1/\tau_0$ can  be expessed in terms of the
%static conductivity $\sigma$  via the Drude formula
%$ \sigma = n e^2 \tau_0 / m$, implying
% $
% 1/ \tau_0 =  \Omega_e^2 / ( 4 \pi \sigma )$.
%The attenuation rate in Eq.~(\ref{eq:Pihydro3}) can then be written expressed in terms of the two phenomenological parameters $\sigma$ and
%$\eta$ as follows
% $1/ \tau_q = \Omega_e^2 /  (4 \pi \sigma) + \frac{4}{3} \eta q^2$.
% 

\section{Summary and conclusions}

In this work we have calculated the energies and the decay rates of the collective charge  modes of ionic crystals in the hydrodynamic regime using the Debye model, 
where the ionic charge density is modeled as a homogeneous, isotropic, elastic medium. 
We have emphasized the crucial role of the direct Coulomb interaction 
$ \mathcal{U}_{ i i } $ 
between ionic charge fluctuations 
for obtaining meaningful results for the collective modes \cite{Hansen23}.
Our main result is that the existence of a sound mode with linear dispersion is immune to long-range Coulomb interactions. 
However, the effective sound velocity depends on the ratio of electronic and ionic plasma frequencies. 
Our conclusion  that in ionic crystals a sound mode with linear dispersion exists even in the presence of long-range Coulomb interactions does not resemble the scenario found in Ref.~\cite{Lucas18}, 
where 
the crossover between electronic sound and plasmons in the hydrodynamic regime of a two-dimensional Fermi liquid was analyzed for a solvable toy model, with the result
that the long-range Coulomb interaction precludes the existence of a sound mode with linear dispersion.

The microscopic calculation of 
the electronic sound velocity $ s $ and the viscosity $ \eta $ entering the hydrodynamic description is beyond the scope of this work.  
In principle this can be done by writing down and solving the relevant kinetic equations, 
but we are not aware of any publications where this has been done for the Debye model. 
According to Ref.~\cite{Huang21}, 
coupled electron-phonon systems are expected to exhibit several regimes characterized by different temperature dependencies of the  parameters entering the hydrodynamic
description.
In particular, 
it would be interesting to investigate whether
the Debye model considered in this work can exhibit a semi-quantum regime \cite{Andreev78,Spivak10},
where $ \eta $ is expected to be proportional to $ 1 / T $ in some temperature range.

\bmhead{Acknowledgements}
	
We thank  Max Hansen and Olexandr Tsyplyatyev  for useful comments and the 
Deutsche Forschungsgemeinschaft (DFG, German Research Foundation) for financial support via  TRR 288 - 422213477.

\section*{Statements and Declarations}

\bmhead{Author contributions}
All authors contributed equally to this work.

\bmhead{Data availability}
Data sharing not applicable to this article as no datasets were generated or analyzed.

\bmhead{Conflict of interest}
The authors have no conflict of interest to declare that is relevant to the content of this article.

\begin{appendix}

\section*{Appendix: Charge correlations and phonon self-energy of the Debye model}

\setcounter{equation}{0}
\renewcommand{\theequation}{A\arabic{equation}}
 \label{sec:ppbubble}

In this appendix
we use functional integral techniques
to derive the equation \eqref{eq:phonselfDebye} relating the phonon self-energy 
$ W ( \bd{q} , \omega )$ 
of the Debye model to its electronic polarization 
$ \Pi_{ e e } ( \bd{q} , \omega ) $.
We also derive the expressions for  the charge correlation functions given in Sec.~\ref{sec:exact}.
Let us start from the Euclidean action of the Debye model,
\begin{align}
S 
= {} & 
\beta \sum_K \left( 
\epsilon_{ \bd{k} } - \mu - i \omega 
\right) \bar{c}_K c_K
\nonumber\\
&
+  \beta \sum_Q \left( 
\omega_{ \bd{q} } - i \bar{\omega} 
\right) \bar{b}_Q b_Q
\nonumber\\
& + \frac{ \beta }{2 \mathcal{V} } \sum_Q 
f_{ \bd{q} } \left( 
\rho^e_{ - Q } - 
\rho^i_{ - Q } 
\right) 
\left( 
\rho^e_{ Q } - 
\rho^i_{ Q }
\right) ,
\label{eq:SDebye}
\end{align}
where $ \beta = 1 /T $ 
is the inverse temperature, 
$ i \omega $ and $ i \bar{\omega} $ 
are fermionic and bosonic Matsubara frequencies, and we have introduced collective labels
$ K = ( \bd{k} , i \omega ) $ and 
$ Q = ( \bd{q} , i \bar{\omega} ) $.
The electrons are now represented by Grassmann fields $ c_K $ and $ \bar{c}_K $ 
while the phonons are represented by a complex field $ b_K $. 
Finally, the electronic and ionic densities are represented as
$ \rho_Q^e  
= \sum_K \bar{c}_K c_{ K + Q} $ and
$ \rho_Q^i 
= - \sqrt{ \mathcal{V} } 
\lambda_{ \bd{q} } X_Q $, 
with 
$ X_Q = 
( b_Q + \bar{b}_{ - Q } ) / 
\sqrt{ 2 \omega_{ \bd{q} } } $ and 
$ \lambda_{ \bd{q} } $ defined in Eq.~\eqref{eq:gammaqdef}.
The Coulomb interaction in the action \eqref{eq:SDebye} can be decoupled via an auxiliary field
$ \phi $ conjugate to the total charge \cite{Kopietz97}. 
The resulting decoupled action is
\begin{align}
S_{cb\phi}  
= {} &
\beta \sum_K \left( 
\epsilon_{ \bd{k} } - \mu - i \omega 
\right) \bar{c}_K c_K
\nonumber\\
&
+ \beta \sum_Q \left( 
\omega_{\bd{q}} - i \bar{\omega} 
\right) \bar{b}_Q b_Q
\nonumber\\
& + \frac{ \beta }{ 2 }  \sum_Q  
f_{ \bd{q} }^{ - 1 } \phi_{ - Q } \phi_Q 
\nonumber\\
&
+ i \frac{ \beta }{ \sqrt{ \mathcal{V} } } 
\sum_Q \left( 
\rho^e_{ - Q } -   
\rho^i_{ - Q } 
\right) \phi_Q .
\label{eq:Scbphi}
\end{align}
Actually, since the ionic density depends only on $ X_{Q} $ 
we may integrate over the canonical phonon momenta to obtain the effective action
\begin{align}
S_{ c X \phi } 
= {} & 
\beta \sum_K \left( 
\epsilon_{ \bd{k} } - \mu - i \omega 
\right) 
\bar{c}_K c_K
\nonumber\\
& +  
\frac{ \beta }{ 2 } \sum_Q 
\Bigl[
\left( 
\bar{ \omega }^2 + \omega^2_{ \bd{q} } 
\right) 
X_{ - Q } X_Q 
+ f_{ \bd{q} }^{ - 1 }  
\phi_{ - Q } \phi_Q 
\nonumber\\
& \hspace{14mm}
+ i \lambda_{ \bd{q} } 
\left( 
X_{ - Q } \phi_Q + 
\phi_{ - Q } X_Q 
\right) 
\Bigr] 
\nonumber\\
& 
+ i 
\frac{ \beta }{\sqrt{ \mathcal{V} } } 
\sum_{ Q , K }   
\bar{c}_{ K + Q } c_{ K } \phi_Q .
\label{eq:ScXphi2}
\end{align}
Note that here, 
the electron-phonon interaction is described by the hybridization 
$ i \lambda_{ \bd{q} } ( 
X_{ - Q } \phi_Q  + 
\phi_{ - Q } X_Q ) $ 
between the phonon displacement and the Coulomb field.
If we are only interested in the electronic properties and in the properties
of the Coulomb-field $ \phi $, 
we may integrate over the phonon field
and arrive at the effective action
\begin{align}
S_{ c \phi } 
= {} & 
\beta \sum_K 
\left( 
\epsilon_{\bd{k}} - \mu - i \omega 
\right) 
\bar{c}_K c_K 
\nonumber\\
& + 
\frac{ \beta}{ 2 } \sum_Q  
\left[ 
f_{ \bd{q} }^{ - 1 } + 
\Pi_{ i i } ( Q ) 
\right] 
\phi_{ - Q } \phi_Q 
\nonumber\\
& + 
i \frac{ \beta }{ \sqrt{ \mathcal{V} } } 
\sum_{ Q , K } 
\bar{c}_{ K + Q } c_{ K } \phi_Q ,
\label{eq:Scphi2}
\end{align}
where
\begin{equation}
\Pi_{ i i } ( Q ) = 
\frac{ \lambda_{ \bd{q} }^2 }{ 
\omega_{ \bd{q} }^2 + \bar{ \omega }^2 
} 
\label{eq:Piii2}
\end{equation} 
can be identified with the imaginary-frequency version of ionic contribution to the interaction-irreducible polarization
introduced in Eq.~\eqref{eq:Piii}.
The total interaction-irreducible polarization 
$ \Pi ( Q ) $ is 
defined by writing the correlation function of the total density in the form
\begin{equation}
\chi ( Q ) =  
\frac{ \Pi ( Q ) }{ 1 + 
f_{ \bd{q} } \Pi ( Q ) }
= 
\frac{ 
\Pi_{ e e } ( Q ) +   
\Pi_{ i i } ( Q ) }{
1 + f_{ \bd{q} } 
\left[  
\Pi_{ e e } ( Q ) + 
\Pi_{ i i } ( Q )  
\right] } ,
\label{eq:chitotDebye}
\end{equation}
where the electronic contribution 
$ \Pi_{ e e } ( Q ) $ 
to the interaction-irreducible polarization is defined via
\begin{equation}
\Pi ( Q ) = 
\Pi_{ e e } ( Q ) + 
\Pi_{ i i } ( Q ) 
= 
\Pi_{ e e } ( Q ) +     
\frac{ \lambda_{ \bd{q} }^2 }{ 
  \omega^2_{ \bd{q} } + \bar{ \omega }^2 
 } .
\label{eq:Pitot}
\end{equation}
With these definitions 
the correlation function of the electronic density can be written as
\begin{align}
\chi_{ e e } ( Q ) 
& =  
\frac{ \Pi_{ e e } ( Q ) }{ 
1 + \frac{ f_{ \bd{q} } }{ 
1 + f_{ \bd{q} } \Pi_{ i i } ( Q ) }  
\Pi_{ e e } ( Q ) }
\nonumber\\
& =  
\frac{ \Pi_{ e e } ( Q )
\left[ 1 + f_{ \bd{q} } \Pi_{ i i } ( Q) \right]
}{  
1 + f_{ \bd{q} }  
\left[  
\Pi_{ e e } ( Q ) + 
\Pi_{ i i } ( Q ) 
\right] } .
\label{eq:chieeim}
\end{align}
Diagrammatically, 
$ \Pi_{ e e } ( Q ) $ 
is irreducible with respect to cutting an effective interaction line representing the partially screened interaction
$ f_{ \bd{q} } / [ 1 
+ f_{ \bd{q} } \Pi_{ i i } ( Q ) ] $.
The real-frequency versions of Eqs.~\eqref{eq:chitotDebye} and \eqref{eq:chieeim}
are given in Eqs.~\eqref{eq:Pidef} and \eqref{eq:chiee} of Sec.~\ref{sec:exact}.

Next, let us derive Eq.~\eqref{eq:phonselfDebye} for the phonon self-energy 
$ W ( \bd{q} , \omega ) $ 
of the Debye model.  
This equation follows from the structure of the generating functional 
$ \Gamma [ X , \phi ] $  
of the vertex functions 
that are irreducible with respect to cutting a single phonon line and a single interaction line. Formally, the functional
$ \Gamma [ X , \phi ] $ 
is defined via a functional Legendre transform of the generating functional 
$ \mathcal{G} [ J^X , J^\phi ] $ 
of the connected correlation functions of the phonon- and Coulomb fields,
which depends on the associated sources 
$ J^X $ and $ J^\phi $ \cite{Kopietz10}. 
Because in the decoupled effective action 
$ S_{ c X \phi } $ 
of the Debye model given in Eq.~\eqref{eq:ScXphi2} 
the phonon field has no self-interaction and does not directly couple to the fermions, 
the generating functional of the
irreducible phonon and Coulomb vertices has the following general form:
\begin{align}
\Gamma [ X , \phi ] 
= {} &
\frac{ \beta }{ 2 } \sum_Q 
\Bigl\{
\left( 
\bar{ \omega }^2 + \omega^2_{ \bd{q} } 
\right) 
X_{ - Q } X_Q
\nonumber\\
& \hspace{1cm}
+ \left[ 
f_{ \bd{q} }^{ - 1 } + 
\Pi_{ e e } ( Q ) 
\right] 
\phi_{ - Q } \phi_Q
\nonumber\\
& \hspace{1cm}
+ i \lambda_{ \bd{q} } 
\left( 
X_{ - Q } \phi_Q + 
\phi_{ - Q } X_Q 
\right) 
\Bigr\} 
\nonumber\\
& + 
\beta 
\sum_{ n = 3 }^{ \infty } \frac{ 1 }{ n ! }
\left( 
\frac{ 1 }{ \sqrt{ \mathcal{V} } } 
\right)^{ n - 2 }
\sum_{ Q_1 \ldots Q_n } 
\nonumber\\
& \phantom{+} \times
\delta_{ 
{ \bd{q} }_1 + \ldots + { \bd{q} }_n , 0 
}
\delta_{ 
\bar{ \omega }_1 + \ldots + \bar{ \omega }_n , 0
}
\nonumber\\
& \phantom{+} \times
\Gamma^{ ( n ) } ( Q_1 , \ldots , Q_n )  
\phi_{ Q_1 } \ldots \phi_{ Q_n } ,
\label{eq:GammacXphi}
\end{align}
where 
$ \Pi_{ e e } ( Q ) $ 
is the interaction-irreducible electronic polarization 
and the 
$ \Gamma^{ ( n ) } ( Q_1 , \ldots , Q_n ) $ 
are the higher-order interaction-irreducible vertices. 
The crucial point is now that in the vertex expansion \eqref{eq:GammacXphi} 
the coefficients of $ X_{ - Q } X_Q $ and of the hybridization vertex $ i \lambda_{\bd{q} } $ are not renormalized, 
because in the bare action \eqref{eq:ScXphi2}  only the Coulomb field couples to the fermions.
This can be easily seen perturbatively and non-perturbatively using the exact hierarchy of functional renormalization group flow equations \cite{Kopietz10}
for the decoupled action \eqref{eq:ScXphi2} of the Debye model. 
The renormalized phonon propagator $ D ( Q ) $ can then be obtained from the upper diagonal element of the propagator matrix
\begin{align}
&
\begin{pmatrix}
\bar{ \omega }^2 + 
\omega_{ \bd{q} }^2 
& 
i \lambda_{ \bd{q} } 
\\
i \lambda_{ \bd{q} }  
& 
f_{ \bd{q} }^{ - 1 } + 
\Pi_{ e e } ( Q )  
\end{pmatrix}^{-1} 
\nonumber\\
= {} &
\frac{ 1 }{  
\left( 
\bar{ \omega }^2 + 
\omega_{ \bd{q} }^2  
\right)
\left[ 
f_{ \bd{q} }^{ - 1 } + 
\Pi_{ e e } ( Q ) 
\right] 
+ \lambda_{ \bd{q} }^2 
}
\nonumber\\
& 
\times
\begin{pmatrix}
f_{ \bd{q} }^{ - 1 } + 
\Pi_{ e e } ( Q )
&
- i \lambda_{ \bd{q} } 
\\
- i \lambda_{ \bd{q} } 
& 
\bar{ \omega }^2 + 
\omega_{ \bd{q} }^2  
\end{pmatrix} ,
\end{align}
implying
\begin{equation}
D ( Q ) = 
\langle X_{ - Q } X_Q \rangle =
\frac{ 1 }{ 
\bar{ \omega }^2 + 
\omega_{ \bd{q} }^2
+ \frac{ 
\lambda_{ \bd{q} }^2 f_{ \bd{q} } 
}{ 
1 + f_{ \bd{q} } \Pi_{ e e } ( Q ) 
}
} .
\label{eq:Dgen}
\end{equation}
After analytic continuation to real frequencies, 
this reduces to Eq.~\eqref{eq:phonselfDebye}.

\end{appendix}


\begin{thebibliography}{99}
%
\bibitem{Anderson63}
P. W. Anderson, {\it{Plasmons, Gauge invariance, and Mass}},
Phys. Rev. {\bf{130}}, 439 (1963)
https://doi.org/10.1103/PhysRev.130.439
%
\bibitem{Schrieffer64}
J. R. Schrieffer, {\it{Theory of Superconductivity}}, (Benjamin/Cummings, Reading, Massachusetts, 1964)
%
\bibitem{Hansen23}
M. O. Hansen, Y. Palan, V. Hahn, M. D. Thomson, K. Warawa, H. G. Roskos, J. Demsar,
F. Pientka, O. Tsyplyatyev, and P. Kopietz,
{\it{Collective modes in the charge density wave state of K$_{0.3}$MoO$_3$: Role of
long-range Coulomb interactions revisited}}, 
Phys. Rev. B {\bf{108}}, 045148 (2023)
https://doi.org/10.1103/PhysRevB.108.045148
%
\bibitem{Pines89}
D. Pines and P. Nozi\`{e}res,
{\it{The Theory of Quantum Liquids Volume I}}, (Addison-Wesley Advanced Book Classics, Redwood City, 1989). See in particular page 172 for the discussion the dielectric function
$\epsilon ( \bd{q} , \omega )$ 
of a charged electron liquid in the hydrodynamic regime. Note that in our notation
the dielectric function is related 
to the electronic irreducible polarization $\Pi_{ee} ( \bd{q} , \omega )$
via $\epsilon ( \bd{q} , \omega ) = 1 + f_{\bd{q}} \Pi_{ee} ( \bd{q} , \omega )$
%
\bibitem{Fetter71}
A. L. Fetter and J. D. Walecka, {\it{Quantum Theory of Many-Particle Systems}},
(McGraw-Hill, New York, 1971)
%
\bibitem{Varga65}
B. B. Varga, {\it{Coupling of Plasmons to Polar Phonons in Degenerate Semiconductors}},
Phys. Rev. {\bf{137}}, A1896 (1965)
https://doi.org/10.1103/PhysRev.137.A1896
%
\bibitem{Mooradian66}
A. Mooradian ad G. B. Wright, {\it{Observation of the Interaction of Plasmons with Longitudinal Optical Phonons in GaAs}}, Phys. Rev. Lett. {\bf{16}}, 999 (1966)
https://doi.org/10.1103/PhysRevLett.16.999
%
\bibitem{Lihm24}
J.-M. Lihm and C.-H. Park,
{\it{Plasmon-Phonon Hybridization in Doped Semiconductors
from First Principles}}, Phys. Rev. Lett. {\bf{133}}, 116402 (2024)
https://doi.org/10.1103/PhysRevLett.133.116402
%
%\bibitem{Fratini08}
%S. Fratini and F. Guinea, {\it{Substrate-limited electron dynamics in graphene}},
%Phys. Rev. B {\bf{77}}, 195415 (2008).
%
%\bibitem{Hwang08}
%E. H. Hwang and S. Das Sarma, {\it{Acoustic phonon scattering limited carrier mobility
%in two-dimensional extrinsic graphene}}, Phys. Rev. B {\bf{77}}, 115449 (2008).
%
\bibitem{Hwang10}
E. H. Hwang, R. Sensarma, and S. Das Sarma, {\it{Plasmon-phonon coupling in graphene}},
Phys. Rev. B {\bf{82}}, 195406 (2010)
https://doi.org/10.1103/PhysRevB.82.195406
%
%\bibitem{Low12}
%T. Low,V. Perebeinos, R. Kim, M. Freitag, and  P.  Avouris, 
%{\it{Cooling of phonoexcited carriers in graphene by internal and substrate phonons}},
%Phys. Rev. B {\bf{86}}, 045413 (2012).
%
\bibitem{Sarma20}
S. Das Sarma and F. Wu, {\it{Electron-phonon and electron-electron interaction effects in twisted
bilayer graphene}}, Ann. Phys. {\bf{417}}, 168193 (2020)
https://doi.org/10.1016/j.aop.2020.168193
%
%\bibitem{Atrazhev95}
%V. M. Atrazhev and I. T. Iakubov, {\it{Viscous dissipation of sound in strongly compressed plasmas}},
%Phys. Plasmas {\bf{2}}, 2624 (1995).
%
\bibitem{Jalalvandi21}
S. Jalalvandi, S. Darbari, and M. K. Moravvej-Farshi,
{\it{Exact dispersion relations for the hybrid plasmon-phonon modes in graphene on dielectric substrates with polar optical
phonons}}, 
Opt. Express {\bf{29}}, 26925 (2021)
https://doi.org/10.1364/OE.434274
%
%
\bibitem{Falter94}
C. Falter and M. Klenner,
\textit{Nonadiabatic and nonlocal electron-phonon interaction and phonon-plasmon mixing in the high-temperature superconductors},
Phys. Rev. B \textbf{50}, 9426 (1994)
https://doi.org/10.1103/PhysRevB.50.9426 
%
%
\bibitem{Falter02}
C. Falter, G. A. Hoffmann, and F. Schnetg\"{o}ke,
\textit{Interlayer phonons and $c$-axis charge response in the high-temperature superconductors},
J. Phys.: Condens. Matter \textbf{14}, 3239
https://doi.org/10.1088/0953-8984/14/12/312
%
%
\bibitem{Bauer09}
T. Bauer and C. Falter,
\textit{Impact of dynamical screening on the phonon dynamics of metallic $La_2 CuO_4$},
Phys. Rev. B \textbf{80}, 094525 (2009)
https://doi.org/10.1103/PhysRevB.80.094525 
%
%
\bibitem{Hepting22}
M. Hepting, M. Bejas, A. Nag, H. Yamase, N. Coppola, D. Betto, C. Falter, M. Garcia-Fernandez, S. Agrestini, Ke-Jin Zhou, M. Minola, C. Sacco, L. Maritato, P. Orgiani, H. I. Wei, K. M. Shen, D. G. Schlom, A. Galdi, A. Greco, and B. Keimer,
\textit{Gapped Collective Charge Excitations and Interlayer Hopping in Cuprate Superconductors},
Phys. Rev. Lett. \textbf{129}, 047001 (2022)
https://doi.org/10.1103/PhysRevLett.129.047001 
%
%
\bibitem{Polini20}
M. Polini and A. K. Geim, {\it{Viscous electron fluids}}, Physics Today {\bf{73}}, 6, 28 (2020)
https://doi.org/10.1063/PT.3.4497
%
%\bibitem{Steinberg58}
%M. S. Steinberg, {\it{Viscosity of the electron gas in  metals}},
%$Phys. Rev. {\bf{109}}, 1486 (1958).
%
%\bibitem{Gurzhi72}
%R. N. Gurzhi and A. I. Kopeliovich, 
%{\it{Electric conductivity of metals with account of phonon drag}},
%Sov. Phys. JETP {\bf{34}}, 1345 (1972).
%
\bibitem{Levchenko20}
A. Levchenko and J. Schmalian, {\it{Transport properties of strongly coupled electron-phonon liquids}},
Ann. Phys. {\bf{419}}, 168218 (2020)
https://doi.org/10.1016/j.aop.2020.168218
%
\bibitem{Huang21}
X. Huang and A. Lucas, {\it{Electron-phonon hydrodynamics}}, Phys. Rev. B {\bf{103}}, 155128 (2021)
https://doi.org/10.1103/PhysRevB.103.155128
%
\bibitem{Kopietz96b}
P. Kopietz, {\it{Bosonization of Coupled Electron-Phonon Systems}}, Z. Phys. B {\bf{100}},  561 (1996)
https://doi.org/10.1007/s002570050162
%
\bibitem{Kopietz97}
P. Kopietz, {\it{Bosonization of Interacting Fermions in Arbitrary Dimensions}}, (Springer, Berlin, 1997)
%
\bibitem{Bohm51}
D. Bohm and T. Staver, {\it{Applications of collective treatment of electron and ion vibrations to theories of conductivity and superconductivity}},
Phys. Rev. {\bf{84}}, 836 (1951)
https://doi.org/10.1103/PhysRev.84.836.2
%
\bibitem{Ashcroft76}
N. W. Ashcroft and N. D. Mermin, {\it{Solid State Physics}} (Holt-Saunders, Philadelphia, 1976)
%
\bibitem{Forster75}
D. Forster, {\it{Hydrodynamic Fluctuations, Broken Symmetry and Correlation Functions}}, (Benjamin/Cummings, Reading, Massachusetts, 1975)
%
%\bibitem{footnotetau}
%In the system where the total momentum is not conserved (for example due to impurities) the damping rate $1/ \tau_q$ in Eq.~(\ref{eq:Pihydro3})
%has the more general form $1/ \tau_q = 1/ \tau_0 + \Gamma q^2 $.
%The momentum-independent part  $1/\tau_0$ can  be expessed in terms of the
%static conductivity $\sigma$  via the Drude formula
%$ \sigma = n e^2 \tau_0 / m$, implying
% $
% 1/ \tau_0 =  \Omega_e^2 / ( 4 \pi \sigma )$.
%The attenuation rate in Eq.~(\ref{eq:Pihydro3}) can then be written expressed in terms of the two phenomenological parameters $\sigma$ and
%$\eta$ as follows
% $1/ \tau_q = \Omega_e^2 /  (4 \pi \sigma) + \frac{4}{3} \eta q^2$.
%
\bibitem{Lucas18}
A. Lucas and S. Das Sarma,
{\it{Electronic sound modes and plasmons in hydrodynamic two-dimensional metals}},
Phys. Rev. B {\bf{97}}, 115449 (2018)
https://doi.org/10.1103/PhysRevB.97.115449
%
\bibitem{Andreev78}
A. Andreev, {\it{Thermodynamics of liquids below the Debye temperature}}, JETP Lett. {\bf{28}}, 557
(1978)
%
\bibitem{Spivak10}
B. Spivak, S. V. Kravchenko, S. A. Kivelson, and X. P. A. Gao,
{\it{Colloquium: Transport in strongly correlated two dimensional electron fluids}}, Rev. Mod. Phys. {\bf{82}}, 1743 (2010)
https://doi.org/10.1103/RevModPhys.82.1743
%
\bibitem{Kopietz10}
See, for example, P. Kopietz, L. Bartosch, and F. Sch\"{u}tz,
{\it{Introduction to the Functional Renormalization Group}} (Springer, Berlin, 2010)
%
\end{thebibliography}
\end{document}